# Mie Resonance Enhancement of Laser Cooling in Rare-Earth Doped Materials

GALINA NEMOVA[1,*] AND CHRISTOPHE CALOZ[1,2]

[1]*Polytechnique Montréal, 2500 ch. de Polytechnique, Montréal, H3T 1J4, QC, Canada.*
[2]*KU Leuven, Kasteelpark Arenberg 10, box 2444, 3001 Leuven, Belgium.*
*\*Corresponding author: [galina.nemova@videotron.ca](galina.nemova@videotron.ca)*

Laser cooling of solids keeps attracting attention owing to abroad range of its applications that extends from *cm*-sized all-optical cryocoolers for airborne and space-based applications to cooling on *nano*particles for biological and mesoscopic physics. Laser cooling of nanoparticles is a challenging task. We propose to use Mie resonances to enhance anti-Stokes fluorescence laser cooling in rare-earth (RE) doped nanoparticles made of low-phonon glasses or crystals. As an example, we consider an $Yb^{3+}$:YAG nanosphere pumped at the long wavelength tail of the $Yb^{3+}$ absorption spectrum at 1030 nm. We show that if the radius of the nanosphere is adjusted to the pump wavelength in such a manner that the pump excites some of its Mie resonant modes, the cooling power density generated in the sample is considerably enhanced and the temperature of the sample is consequently considerably (~ 63%) decreased. This concept can be extended to nanoparticles of different shapes and made from different low-phonon RE doped materials suitable for laser cooling by anti-Stokes fluorescence.

## I. INTRODUCTION

In 1929, Peter Pringsheim proposed the idea to cool sodium vapor with light using anti-Stokes fluorescence [1]. This idea was based on experimental evidence that some materials emit light at shorter wavelength than the material illumination wavelength. The source of this quantum defect was inelastic collisions of sodium atoms that is thermalization. Light cooling in solids by anti-Stokes fluorescence was realized only in 1995, by Richard Epstein and colleagues [2]. They experimentally demonstrated a drop of the temperature of a *cm*-sized ytterbium-doped fluorozirconate $ZrF_4$-$BaF_2$-$LaF_3$-$AlF_3$-$NaF$-$PbF_2$ (ZBLANP) glass sample by 0.3 K below room temperature. This success was largely owed to the recent development of high-quality ZBLANP for the telecommunication industry and of coherent light sources (lasers).

Since its first experimental demonstration by Epstein, laser cooling of solids via anti-Stokes fluorescence has been realized with different rare-earth (RE) ions, including $Yb^{3+}$, $Er^{3+}$, $Tm^{3+}$ and $Ho^{3+}$ ions, doped in a wide variety of low-phonon glasses and crystals [3–8]. The ytterbium ($Yb^{3+}$) ions have eventually been found to represent the best choice for laser cooling applications. Indeed, ytterbium has only one excited manifold, and it therefore is free from excited state absorption, which is a source of undesirable non-radiative decay that results in heat generation. A record low temperature of $\sim 91\,K$ has been achieved in an $Yb^{3+}$:YLF sample [9, 10].

The technology of laser cooling by anti-Stokes fluorescence has numerous applications. For example, it can be used for development a solid-state all-optical cryocoolers, which are based on *cm*-sized RE-doped low-phonon crystal or glass samples. Such all-optical cryocoolers are compact, reliable, and free from vibrations and moving parts or fluids devices. They can be used for the cryogenic refrigeration of semiconductor devices in applications with tight weight and size constraints, such as airborne and space-based detector systems. The first all-solid-state optical cryocooler was demonstrated in 2018 [11].

Laser cooling by anti-Stokes fluorescence can also be realized in radiation-balanced solid-state bulk lasers where the heat generated by the quantum defect between the pump and laser signals is offset by cooling from anti-Stokes emission [12]. This technology can be used for mitigating heat generation in disc lasers [13, 14]. Specifically, researchers proposed the idea to cool an RE-doped fiber laser cladding with anti-Stokes fluorescence to mitigate heat generated in the RE-doped fiber laser core during the amplification process [15], this proposal was experimentally demonstrated in [16, 17].

The size and shape of the laser cooled sample as well as RE ion concentration in this sample are important parameters in the optimization of the laser cooling process [18]. In the resent years, interest has emerged in the laser cooling of small, *μm*- and *nm*-sized RE-doped samples [19-21]. Such small samples are of interest for biological applications and mesoscopic physics. However, laser cooling at the nanoscale is not an easy task [18].

In this paper, we propose an approach to enhance the laser cooling process in RE–doped nanospheres with a diameter of a few hundreds of nanometers. Our approach is based on optimization of the sample size by leveraging the Mie resonance concept and theory [22]. As an example, an $Yb^{3+}$:YAG nanosphere pumped at the wavelength $\lambda_P = 1030\,nm$ located in the long wavelength tail of the $Yb^{3+}$ absorption spectrum is considered (Fig. 1). The mean fluorescence wavelength of the spontaneous radiation $\lambda_F < \lambda_P$. A bulk $Yb^{3+}$:YAG sample was laser cooled in [23]. Section II describes the theory of laser cooling in bulk and nanosphere samples. Section III presents and discusses the results. Conclusions are given in Sec. IV.

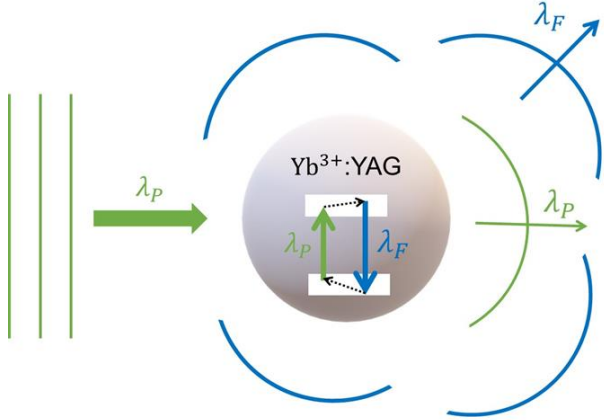

Fig. 1 Nanosphere system under investigation. Here $\lambda_P$ and is $\lambda_F$ are the pump and mean fluorescence wavelengths, respectively.

## II. THEORETICAL ANALYSIS

### A. Laser Cooling in Yb$^{3+}$:YAG

The process of laser cooling with anti-Stokes fluorescence of RE-doped samples has been considered in details in a number of works [6, 8, 18]. Here we only briefly discuss this process. It is well known that the energy levels of RE ions doped in a host material undergo splitting as a result of the Stark effect. In the case of Yb$^{3+}$:YAG samples two energy levels $^2F_{7/2}$ and $^2F_{5/2}$ of Yb$^{3+}$ ions split into four and three sublevels, respectively (Fig. 2a). It is worth to remind that YAG (yttrium aluminum garnet) is one of the most widely used optical materials in laser physics and laser cooling of solids, due to its high thermal stability, good chemical resistance, and low-phonon energy. In thermal equilibrium, the population of each sublevel is described by the Boltzmann distribution. This thermal equilibrium can be reached via phonon absorption and phonon emission. Indeed, each electron at each sublevel of the ions interact with phonons of the host material by absorbing phonons at the rate $A_{NR}^+$ or emitting phonons at the rate $A_{NR}^-$, as shown in Fig. 2a. This process, known as the thermalization, takes place at the *ps*-time scale.

Let us consider a laser cooling cycle in Yb$^{3+}$:YAG. It includes phonon absorption at the wavelength $\lambda_P$, thermalization in the exited $^2F_{5/2}$ manifold accompanied by phonon absorption, and spontaneous photon emission from the exited $^2F_{5/2}$ manifold to the ground $^2F_{7/2}$ manifold. All photon absorption-emission cycles where the energy of the radiated photon exceeds the energy of the absorbed one (anti-Stokes fluorescence) result in cooling, since this energy difference has to be compensated by phonon absorption. All the photon absorption-emission cycles where the energy of the absorbed photons exceeds the energy of the radiated ones (Stokes fluorescence) causes heat generation in the sample,

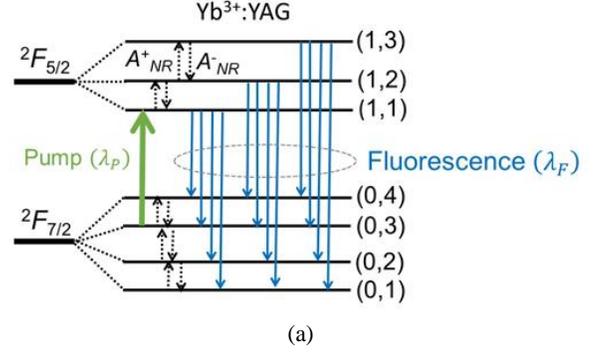

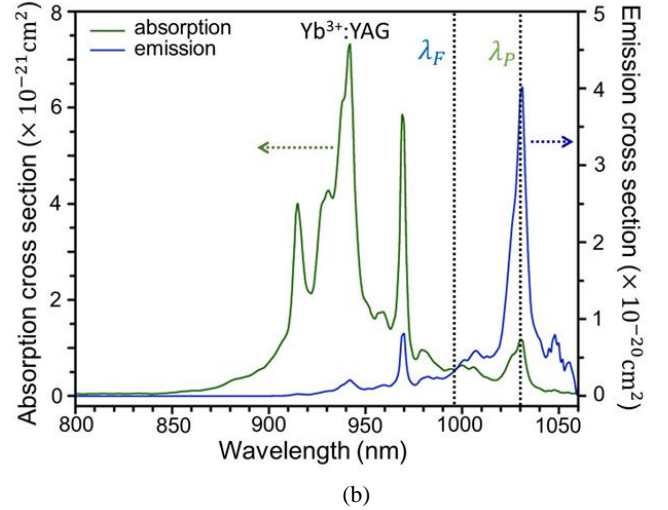

Fig. 2 Spectroscopy of the Yb$^{3+}$:YAG sample. (a) Energy levels. (b) Absorption and emission cross sections versus wavelength.

since the energy difference in these cycles has to be compensated by phonon generation (Fig. 2a).

Spontaneous emission is characterized by the mean fluorescence frequency

$$\nu_F = \frac{\int \Phi(\nu)\nu d\nu}{\int \Phi(\nu)d\nu}, \qquad (1)$$

where $\Phi(\nu)$ is the fluorescence flux density, which can be obtained experimentally, and $\nu_F = c/\lambda_F$. Here $\lambda_F$ is the mean fluorescence wavelength and $c$ is the speed of light in vacuum. If one wants to cool a sample, the pump wavelength, $\lambda_P$, has to exceed the mean fluorescence wavelength, $\lambda_F$, and all radiated photons with wavelengths smaller then $\lambda_F$ must leave the sample (not to be reabsorbed by Yb$^{3+}$ ions). Indeed, these short-wavelength reabsorbed photons serve as the source of heat generation in the sample, which is undesirable for cooling. The efficiency of the cooling cycle can be estimated as the difference between the energies of the anti-Stokes emitted photon and the pump photon normalized by the energy of the pump photon that is

$$\eta_{cool} = \frac{h\nu_F - h\nu_P}{h\nu_P} = \frac{\lambda_P}{\lambda_F} - 1. \qquad (2)$$

The electrons excited to the $^2F_{5/2}$ manifold can also decay nonradiatively to the ground $^2F_{7/2}$ manifold generating phonons. The "competition" between the radiative and nonradiative decays of the $^2F_{5/2}$ electrons can be quantified by the quantum efficiency

$$\eta_q = \frac{W_r}{W_r + W_{nr}} \quad (3)$$

where $W_r = 1/\tau_r$ and $W_{nr} = 1/\tau_{nr}$ are the radiative and nonradiative decay rates, respectively, with $\tau_r$ and $\tau_{nr}$ being the radiative and nonradiative lifetimes of the excited manifold, respectively. We consider only high-quality Yb$^{3+}$:YAG samples with the quantum efficiency as high as $\eta_q \geq 99\%$. Only such samples are suitable for laser cooling. Indeed, in this case the probability of nonradiative decay with phonon generation from the exited $^2F_{5/2}$ to the ground $^2F_{7/2}$ manifold is very low [8].

The strength of the pump absorption process can be quantified by the absorption cross section (Fig. 2b), which can be obtained experimentally. For example, if the intensity of the pump source at the wavelength $\lambda_P$ is $I_P(\lambda_P)$, the number of photons absorbed per unit volume of the RE-doped sample per second is $N_T \sigma_a(\lambda_P) I_P(\lambda_P)/(h\nu_P)$, where $N_T$ is the RE-ion concentration in the sample, $h$ is the Planck constant, $\nu_P = c/\lambda_P$ is the pump photon frequency. The cooling power generated in the sample, $P_{cool}$, can be estimated as the difference between the absorbed and radiated powers. In the steady-state regime the cooling power generated in the sample with the volume $V$ is [6]

$$P_{cool} = \frac{h\tilde{\nu}_F - h\nu_P}{h\nu_P} \frac{V N_T \sigma_a(\nu_P) I_P}{\left[1 + \frac{I_P}{I_S}\left(1 + \frac{\sigma_e(\nu_P)}{\sigma_a(\nu_P)}\right)\right]} \approx$$

$$\approx \frac{h\tilde{\nu}_F - h\nu_P}{h\nu_P} V N_T \sigma_a(\nu_P) I_P, \quad (4)$$

where $I_S = h\nu_P/[\eta_q \tau_r \sigma_a(\nu_P)]$ is the saturation intensity and $\tilde{\nu}_F = (2\eta_q - 1)\nu_F \approx \eta_q \nu_F$ is the effective mean fluorescence frequency. If the sample is placed on a low-contact and low-thermal conductivity support in a vacuum chamber, only radiative heat load may take place in the system. The final equilibrium temperature of the sample, $T_S$, can then be estimated by the Stefan-Boltzmann law,

$$P_{cool} = \varepsilon \sigma_B S (T_r^4 - T_S^4), \quad (5)$$

where $S$ is the total surface area of the sample, $T_r$ is room temperature, and $\varepsilon$ is the sample emissivity. As one can understood from (2), the efficiency of laser cooling cycles is very low (a few %). Enhancement of the laser cooling process is therefore highly desirable.

### B. Mie Resonance Cooling

Les consider the laser cooling process in a Yb$^{3+}$:YAG nanosphere pumped with a plane electromagnetic wave propagating vacuum at the pump wavelength $\lambda_P$ (Fig. 1). According to Mie theory, the pump plane wave can be expanded in spherical harmonics as

$$\vec{E}_P = E_0 \sum_{n=1}^{\infty} i^n \frac{2n+1}{n(n+1)} \left[\vec{M}_{o1n}^{(1)} - i\vec{N}_{e1n}^{(1)}\right], \quad (6)$$

where $E_0$ is the amplitude of the incident field, $\vec{M}_{o1n}$ and $\vec{N}_{e1n}$ are the vector spherical harmonics

$$\vec{M}_{o1n}^{(1)} = \cos\varphi \cdot \pi_n(\cos\theta) \cdot j_n(\tilde{r})\hat{e}_\theta - \sin\varphi \cdot \tau_n(\cos\theta) j_n(\tilde{r})\hat{e}_\varphi, \quad (7)$$

$$\vec{N}_{e1n}^{(1)} = n(n+1)\cos\varphi \cdot \sin\theta \cdot \pi_n(\cos\theta) \cdot \frac{j_n(\tilde{r})}{\tilde{r}}\hat{e}_r + \cos\varphi \cdot \tau_n(\cos\theta) \frac{[\tilde{r}j_n(\tilde{r})]'}{\tilde{r}}\hat{e}_\theta - \sin\varphi \cdot \pi_n(\cos\theta) \frac{[\tilde{r}j_n(\tilde{r})]'}{\tilde{r}}\hat{e}_\varphi,$$

where $\tilde{r} = 2\pi n_{Yb} r/\lambda_P$ and $r$ is the radial distance. The index $n = 1$ stands for dipole, $n = 2$ quadrupole, etc., $j_n$ are the spherical Bessel functions, and $\pi_n$ and $\tau_n$ are the angle-dependent functions defined as

$$\pi_n = \frac{P_n^1}{\sin\theta}, \text{ and } \tau_n = \frac{dP_n^1}{d\theta}, \quad (8)$$

where $P_n^1$ in the associated Legendre function of first kind of degree $n$ and the first order. The field inside the sphere reads [22, 24]

$$\vec{E} = E_0 \sum_{n=1}^{\infty} i^n \frac{2n+1}{n(n+1)} \left[c_n \vec{M}_{o1n}^{(1)} - id_n \vec{N}_{e1n}^{(1)}\right], \quad (9)$$

where the coefficients $c_n$ and $d_n$ are functions of the normalized radius of the sphere, $x = 2\pi R/\lambda_P$, with $R$ being the radius of the sphere, and read

$$c_n = \frac{j_n(x)\left[xh_n^{(1)}(x)\right]' - h_n^{(1)}(x)[xj_n(x)]'}{j_n(mx)\left[xh_n^{(1)}(x)\right]' - h_n^{(1)}(x)[mxj_n(mx)]'}, \quad (10)$$

$$d_n = \frac{mj_n(x)\left[xh_n^{(1)}(x)\right]' - mh_n^{(1)}(x)[xj_n(x)]'}{m^2 j_n(mx)\left[xh_n^{(1)}(x)\right]' - h_n^{(1)}(x)[mxj_n(mx)]'}, \quad (11)$$

where $h_n^{(1)}$ is the spherical Hankel functions, $m = n_{Yb}$. The $c_n$ coefficients describe the magnetic multipole modes and the $d_n$ coefficients describe the electric multipole modes. At values of $R$ for which the denominator of $c_n$ coefficient or the denominator of $d_n$ vanishes, the sphere resonantes and scattering exhibits therefore a peak that dominates the response of the sphere. The radial dependence of the absolute-square of the electric field averaged over polar angle, $\theta$, and azimuthal angle, $\varphi$, of spherical coordinates has the form

$$\langle |\vec{E}|^2 \rangle = \frac{E_0^2}{4} \sum_{n=1}^{\infty} (m_n |c_n|^2 + n_n |d_n|^2), \quad (12)$$

where $m_n$ and $n_n$ are functions of the radial distance, $r$, and are given as

$$m_n = 2(2n+1)|j_n(\tilde{r})|^2, \quad (13)$$

$$n_n = 2n(2n+1)\left[(n+1)\left|\frac{j_n(\tilde{r})}{\tilde{r}}\right|^2 + \left|\frac{(\tilde{r}\cdot j_n(\tilde{r}))'}{\tilde{r}}\right|^2\right], \quad (14)$$

where $\tilde{r} = 2\pi n_{Yb} r/\lambda_P$.

The cooling power generated in the sphere can be estimated as the difference between the pump power absorbed by the $Yb^{3+}$ ions in the sphere and the power radiated with anti-Stokes fluorescence at the mean fluorescence wavelength. The field intensity given by (12) can then be integrated over the nanosphere and then inserted into (4) as

$$P_{cool} = 4\pi\left(\frac{h\tilde{\nu}_F - h\nu_P}{h\nu_P}\right) N_T \sigma_a(\nu_P)\sqrt{\frac{\varepsilon_0}{\mu_0}} \int_0^R \langle|\vec{E}|^2\rangle r^2 dr, \quad (15)$$

which corresponds to the cooling power generated in the Mie-resonant nanosphere.

## III. RESULTS AND DISCUSSION

Let us consider an RE-doped nanosphere with radius of a few hundred nanometers. The energy levels of $Yb^{3+}$ ions in a bulk $Yb^{3+}$:YAG sample, shown in Fig. 2a, remain essentially unchanged when the size of the sample is reduced to a few hundred nanometers. Moreover, phonon quantization, which can affect the thermalization process in the RE-doped nanocrystal, is inessential in such samples. We consider, as an example, a sample with ion concentration of ~10%, which is smaller than the critical concentration of $Yb^{3+}$:YAG (17%). Such samples are free from cooperative effects such as resonant-radiative transfer (reabsorption), resonant nonradiative transfer and so on, which deteriorate laser cooling [6]. Indeed, these cooperative effects result to the excited energy migration from the ion to ion inside the sample. This migrating energy can excite impurities in the sample and decay nonradiatively causing heat generation in the sample. The pump power considered in all our simulations is $I_P = 0.5 \cdot 10^{-4}\ W/\mu m^2$. Physical properties and parameters of $Yb^{3+}$:YAG samples used in our simulations are summarized in Table 1.

As one can see in relations (5), the temperature of the sample depends on the cooling power generated in the sample. The cooling power generated in the nanosphere can be estimated with relation (15), which depends on the electric field intensity distribution, $\langle|\vec{E}|^2\rangle$, inside the sample. This distribution depends on the Mie resonant modes. Let us consider Mie modes of spheres with different radii. As may be seen in (10, 11), if the $c_n$ coefficient exhibits a peak, a magnetic multipole mode dominates the response of the sphere, while if the $d_n$ coefficient exhibits a peak, an electric multipole mode dominates the sphere. These coefficients exhibit peaks when their denominators tend to zero. For each $n$, there are several radii for which the denominators of $c_n$ and $d_n$ tend to zero. For example, at the pump wavelength $\lambda_P = 1030\ nm$ the magnetic dipole (MD) mode ($n = 1$) can be supported by the sphere with the resonant radius $\sim 200\ nm$ or $\sim 485\ nm$, the magnetic quadrupole (MQ) modes ($n = 2$) can be supported by the sphere with the resonant radius $\sim 312\ nm$ or $\sim 600\ nm$. The electric dipole (ED) mode ($n = 1$) can be supported by the sphere with the resonant radius $\sim 325\ nm$; and the electric quadrupole (EQ) mode ($n = 2$) can be supported by the sphere with the resonant radius $\sim 423\ nm$. The absolute-square of the electric field averaged over polar angle, $\theta$, and azimuthal angle, $\varphi$, of spherical coordinates, $\langle|\vec{E}|^2\rangle$, for the ED mode of the sphere with the resonant radius $\sim 325\ nm$ and for the MQ mode of the sphere with the resonant radius $\sim 312\ nm$ is presented in Fig. 3. These field distributions have been simulated with relation (12).

Table 1. Physical properties of YAG and $Yb^{3+}$:YAG

| YAG | |
|---|---|
| Crystal structure | cubic |
| Density (g·cm$^{-3}$) | 4.56 |
| Thermal expansion (C$^{-1}$) | 7.8×10$^{-6}$ |
| Thermal conductivity (W·m$^{-1}$·C$^{-1}$) | 13 |
| Refractive index | 1.83 |
| Phonon energy (cm$^{-1}$) | ~630 |
| $Yb^{3+}$:YAG | |
| $\lambda_P$ (nm) | 1030 |
| $\sigma_a(\lambda_P)$ (cm$^2$) | 1.25×10$^{-21}$ |
| $\sigma_e(\lambda_P)$ (cm$^2$) | 4×10$^{-20}$ |
| $\tau_r$ (μs) | ~951 |
| $N_T$ | 10% |
| Critical $Yb^{3+}$ concentration (ions/cm$^{-3}$) | 2.3×10$^{21}$ |

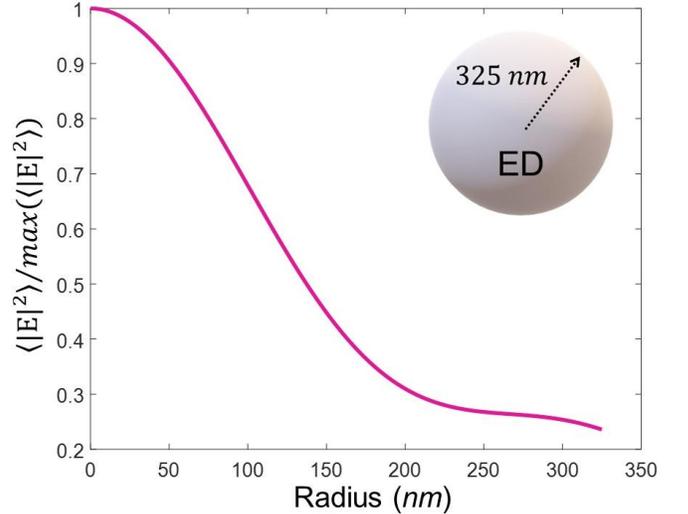

(a)

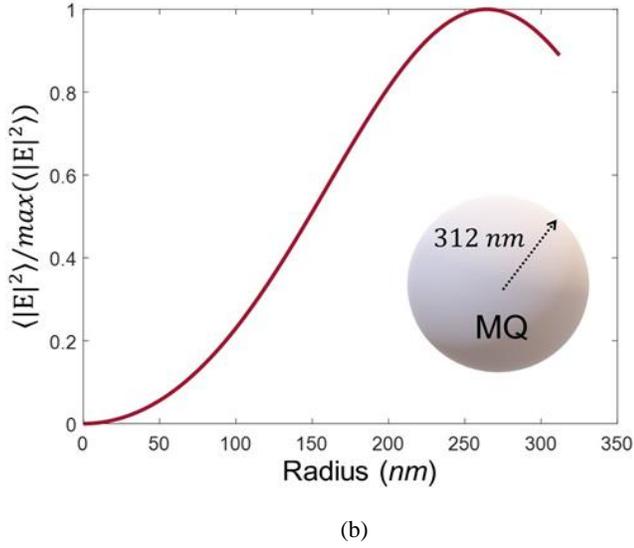

(b)

Fig. 3. Dispribution of the squared modulus of the electric field averaged over the polar angle, $\theta$, and the azimuthal angle, $\varphi$, along the radius of the sphere. (a) Electric dipole mode in the sphere with the radius $R \approx 325\ nm$. (b) Magnetic quadrupole mode in the sphere with the radius $R \approx 312\ nm$. These results have been computed using (12).

The cooling power density generated in the nanosphere, $\rho_{cool} = P_{cool}/V$, where $V = \frac{4}{3}\pi R^3$ is the volume of the sphere, can be estimated using (15). $R$ is the radius of the sphere. The cooling power density for nanospheres with different radii in the range between $150\ nm$ and $650\ nm$ is plotted in Fig. 4. It changes considerably on this range and has several peaks. The radii of the nanospheres corresponding to these peaks are very close to the resonant radii of the Mie resonant modes.

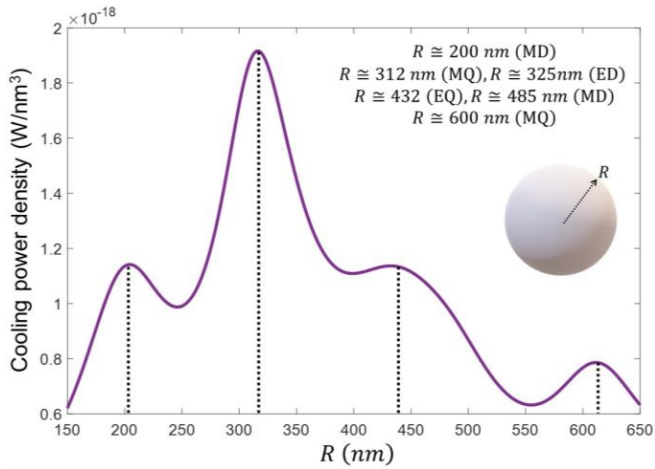

Fig. 4. Cooling power density generated in nanospheres with different radii. Resonant radii of the nanospheres supporting MD, ED, MQ, and EQ mode are presented for simplity of understanding.

As may be seen in Fig. 4, the values of these peaks are different from each other. The largest peak corresponds to the radius $\sim 317\ nm$. The nanosphere with this radius can support both the MQ and ED modes. The resonant radii for these modes ($\sim 312\ nm$ and $\sim 325\ nm$) are very close to the radius of the nanosphere $\sim 317\ nm$. If the radius of nanosphere is $\sim 200\ nm$, only the MD mode can be supported by the sample. If the radius of the nanosphere is equal to $\sim 430\ nm$ the peak is very broad, since one of the Mie resonant modes (the EQ mode) has the resonant radius, which is very close the radius of this peak, but the MD mode has the resonant radius $\sim 485\ nm$, which is essentially different from the radius of the nanosphere ($\sim 430\ nm$). This MD mode only weekly influence this peak. The peak corresponding to the nanosphere with the radius $\sim 612\ nm$ is small. It is located near the resonant radius of the MQ mode, which is $\sim 600\ nm$. There are no any other Mie modes with the resonant radii close to $\sim 612\ nm$.

With the knowledge of the cooling power density, it is possible to estimate the temperature of the nanospheres using relation (5). These results are presented in Fig. 5, which plots the temperature of the nanosphere, $T_S$, versus its radius in the same radius range between $150\ nm$ and $650\ nm$ as the cooling power density has been calculated in Fig. 4.

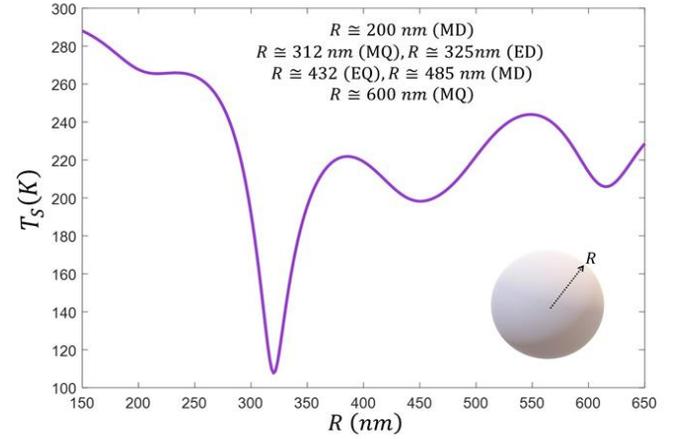

Fig. 5. Sample temperature, $T_S$, versus the radius of the sample. Resonant radii of the nanospheres supporting MD, ED, MQ, and EQ mode are presented for simplity of understanding.

As one might expect, the radii of the nanospheres for which temperature reaches minima almost perfectly coincide with those associated with highest cooling power density (Fig. 4), the small discrepancy between the two being caused by the nonlinear relation between the cooling power and the temperature of the nanosphere (5), which for the spherical sample has the form $R\rho_{cool} = 3\varepsilon\sigma_B(T_r^4 - T_S^4)$. As is obvious from this relation, the radius of the nanosphere and the cooling power density generated in the nanosphere influence the drop in the temperature of the sample. The cooling power density generated in the nanospheres with the radii $\sim 200\ nm$ and $\sim 432\ nm$ are almost equal to each other (Fig. 4), but the temperature of the nanosphere with the radius $\sim 432\ nm$ exceeds the temperature of the nanosphere with the radius $\sim 200\ nm$ (Fig. 5). The minimal temperature in Fig. 5 is $T_S = 107\ K$. It can be reached if the radius of the nanosphere is equal to $\sim 320\ nm$. The temperature of this nanosphere drops by ~ 60% of the temperature of the nanosphere ($T_S = 265\ K$) with the radius ~ $250\ nm$; it drops by ~ 63% of the temperature of the nanosphere ($T_S = 288\ K$) with the radius ~ $150\ nm$, and by ~ 51% of the temperature of the nanosphere ($T_S = 222\ K$) with the radius ~ $375\ nm$. This considerable decrease in the temperature of the nanosphere at the radius $R = 320\ nm$ is caused by the enhancement of the cooling power density in the sample which is itself caused by the ED and MQ Mie resonating modes.

## IV. CONCLUSIONS

We have proposed to use Mie resonant modes to enhance laser cooling in RE-doped nanospheres made of low-phonon glasses or crystals. This concept have been comprehensively investigated for $Yb^{3+}$:YAG nanospheres with radii in the order of a few hundred nanometers. We have considered $Yb^{3+}$:YAG nanospheres pumped in the long-wavelength tail of the $Yb^{3+}$ absorption spectrum at the wavelength $\lambda_P = 1030\ nm$. We have shown a considerable enhancement of the cooling power density due to the Mie resonant modes and a related substantial (~ 60%) decrease in the temperature of the laser cooled samples with the radius ~ $320\ nm$. This sample supports MQ and ED Mie modes. A smaller decrease in the temperature of the laser cooled nanosphere is obtained if the sample supports only one Mie mode, as in the case of the $Yb^{3+}$:YAG nanospheres with the radii ~ $200\ nm$, ~ $450\ nm$, and ~ $650\ nm$. Nanospheres with the radii hundred nanometers are used widely in biophysics and mesoscopic physics.

As we already mentioned for samples of such a size scale, phonon quantization, which can influence the thermalization process, the change in the energy levels of the RE ions doped in the sample, as well as the Purcell effect, are not essential. Quantization becomes important only for spheres with smaller radii, in the order of a few nanometers. For such small sphere, quantization can dramatically enhance the laser cooling process, but the related technology becomes naturally challenging.

A lot of theoretical and experimental research is currently devoted to the laser cooling of nanoparticles. In the majority of the related work, only the translational energy of the particles has been suppressed. Removal of the internal energy embedded in the vibrational degrees of freedom of the particles, which has been considered in this work, remains an unsettled problem. Nowadays, laser cooling with anti-Stokes fluorescence is the only possible technique permitting to remove the internal energy of the sample, but the efficiency of this technique is very low (a few %). Our proposed approach based on the Mie resonant modes may therefore be considered as a timely new tool for experimentalists working on the laser cooling of solid nano-samples.